# Classical Lagrange Formalism for Non-Conservative Dynamical Systems

**Alex Ushveridze**

Algostream Consulting, Minneapolis, MN, USA

alexush@algostream.com

## *Abstract*

*The classical Lagrange formalism is generalized to the case of arbitrary stationary (but not necessarily conservative) dynamical systems. It is shown that the equations of motion for such systems can be derived in the standard ways from the Lagrange functions which: i) are linear in system velocities, ii) are stationary, i.e., have no explicit dependence on time, and iii) do not require the introduction of any additional degrees of freedom. We show that applying Noether theorem to such Lagrange functions leads to the integrals of motion naturally generalizing the notion of energy but not coinciding with it in non-physical and/or non-conservative cases. The non-conservative analogs of Hamilton equations, Poisson brackets, Hamilton-Jacobi equations, Liouville theorem and Principle of Stationary Action are discussed as well. As an example, we consider two cases demonstrating the work of the proposed schema: i) the 'classical' model of one-dimensional damped harmonic oscillator and ii) a special class of non-physical exactly-solvable multi-dimensional dynamical models.*

## 0. Introduction

### 0.1. The problem

It is generally accepted that the equations of motion for non-conservative[1] physical systems cannot be derived from the stationary[2] Lagrange functions within the standard variation approach. The main argument in favor of this point of view is based on the use of Noether theorem, according to which the stationarity of a given Lagrangian inevitably leads to the existence of a corresponding integral of motion. If, following the standards of theoretical physics, one identifies such an integral with energy, then one immediately arrives at the energy conservation law for the system in question. This means that the latter cannot be non-conservative [1-2].

However, who said that time symmetry-based integral of motion must necessarily be related to the energy of the system? Assume that we do not know anything about physics and want to consider the problem exclusively from mathematical point of view. If so, why not associate the Noetherian integral with one of the $N-1$ independent integrals of motion, whose existence for any N-dimensional systems can be guaranteed with no regard to whether the system is conservative or not? Could this naturally solve the "inverse Lagrange problem" for non-conservative (and maybe even non-physical) dynamical equations?

---

[1] The adjective 'non-conservative' will refer to both non-physical systems (i.e., systems for which the notion of energy cannot be defined in meaningful ways) and dissipative physical systems (for which the energy is definable, but its conservation cannot be guaranteed). The simplest example of the latter systems is the 'classical' model of damped harmonic oscillator.

[2] The adjective 'stationary' will refer to systems with no explicit dependence on time.

These are the questions we plan to address in this paper. Note that the inverse Lagrange problem for stationary dissipative systems has a long history, beginning with the seminal paper [3]. There is a surprisingly large number of publications (even the recent ones -- see e.g., the detailed review [4] and many references therein) discussing various aspects of this problem and proposing diverse methods of its solution. The latter include both the extensions of the 'classical' Lagrange formalism [5] and some exotic solutions such as the time-dependent Lagrangians [6,7], Lagrangians with extra degrees of freedom [8], approximate solutions [9], etc.). At the same time, we are aware only a few papers (like [10-12]) in which the idea of linking the Lagrangian to non-traditional integrals of motion was articulated explicitly. The authors of [13] have illustrated this idea with the example of one-dimensional damped harmonic oscillator.

In this paper, we plan to consider the inverse Lagrangian problem for the most general case of dynamical equations. There will be no specific restrictions on their shape and dimension: we allow these equations to be completely "non-physical". In other words, we're not going to associate any physical concepts like "energy," "momentum," "coordinate," etc. with these equations. Despite that, we will still use many of such "physical" terms in this article, in order to emphasize similarity of mathematical constructions we plan to discuss here with those used in the physical literature. We will show that practically all the basic mathematical structures appearing in the rich formalism initially developed for describing the stationary and conservative systems of classical mechanics can naturally be generalized to the case of dynamical equations of the most general type (including the strictly 'non-physical' ones). Below we will present a general mathematical framework for such a generalization, focusing primarily on its mathematical and structural aspects.

The questions discussed in this paper could be interesting to two potential audiences: 1) theoretical physicists still believing that no Lagrange formulation of dissipative systems is possible -- they may be surprised to learn that this is not quite so, and 2) data scientists interested in developing the ML/AI algorithms for converting the sequences of raw observations into predictive models – they could be intrigued by the fact that the models they build (being essentially represented by dynamical systems of the most general type) allow multiple equivalent descriptions and thus can be approached within different algorithmic methodologies. The latter audience is our primary target. This may explain the style of our exposition (which some may find too overloaded with details). Those who prefer to get to the point quickly can jump directly to its last Section 8, in which all the basic results are listed in a fairly dense form accompanied with the references to all the relevant places in the main body of text.

### 0.2. The equations we are going to 'Lagragianize'

Our primary focus in this paper will be the system of $N$-dimensional first-order dynamical equations

(0.1)  $\dot{x}_i = R_i(x), \quad i = 1, \ldots, N$

in which $x = \{x_1, \ldots, x_N\}$ denotes the set of $N$ dynamical variables (coordinates).

The fact that we are going to start with the first-order equations should not lead to any loss of generality because any system of dynamical equations of a finite order can be reduced to the first-order form by considering the derivatives of dynamical variables (the velocities, and, if needed, their higher derivatives) as new independent variables. For example, the second-order Newtonian equation for the damped harmonic oscillator $m\ddot{x} + \gamma\dot{x} + kx = 0$, after redefinitions $x_1 = \dot{x}$, $x_2 = x$, can be rewritten as a system of two first-order equations: $\dot{x}_1 = x_2$, $\dot{x}_2 = (\gamma/m)x_1 + (k/m)x_2$.

The functions in the right-hand side of (0.1) are assumed to be arbitrary sufficiently smooth functions of coordinates $x$. We are not imposing any additional constraints on their form and are not requiring any special symmetries for them except their invariance under time-translations. This means that we are limiting ourselves to the stationary systems only (i.e., those whose right-hand sides do not explicitly depend on time).

Note that the condition of stationarity of a system does not imply its conservativity. For example, the abovementioned damped harmonic oscillator is stationary if $m, k$ and $\gamma$ do not depend on time but it is non-conservative if $\gamma \neq 0$ (provided that the energy $E = (m\dot{x}^2 + kx^2)/2$ is still defined as in conservative case with $\gamma = 0$).

It is worth stressing again that the equations (0.1) are not expected to be 'physical' at all, so it could be impossible to associate with them the notion of physical energy in any meaningful ways. In that case characterizing the equations of the type (0.1) as 'conservative' or 'non-conservative' is not fully correct.

## 0.3. The conservative systems – a particular case

The particular case of conservative (and fully 'physical') case arises from (0.1) when the entire family of $N$ functions $R_i(x)$ is parametrized by a single function of dynamical variables $H(x)$ as:

$$(0.2) \qquad R_i(x) = \sum_{k=1}^{N} \sigma_{ik} \frac{\partial H(x)}{\partial x_k}, \quad i = 1, \dots, N$$

Here $\sigma_{ik}$ is an arbitrary *constant and invertible anti-symmetric tensor*. The function $H(x)$ – which is called the Hamiltonian is what one typically associates with the energy of the system. In the stationary case which we are limiting to here, the Hamiltonian $H(x)$ is a natural integral of motion, so the energy is conserved. This can be proved by directly calculating its full time-derivative $\dot{H}(x)$ by applying the chain rule to the right-hand side (0.2) of the equation (0.1).

Such systems allow a very rich and elegant mathematical description within both the Hamilton and Lagrange formalisms [2]. The corresponding Hamilton equations have the form

$$(0.3) \qquad \dot{x}_i = \sum_{k=1}^{N} \sigma_{ik} \frac{\partial H(x)}{\partial x_k}, \quad i = 1, \dots, N$$

and the corresponding Lagrange function is given by

$$(0.4) \qquad L(x, \dot{x}) = \sum_{k=1}^{N} P_k(x)\dot{x}_k - H(x)$$

where

$$(0.5) \quad P_k(x) = \frac{1}{2}\sum_{i=1}^{N} \sigma_{ik}^{-1} x_i$$

and $\sigma_{ik}^{-1}$ is another anti-symmetric tensor -- the inverse of $\sigma_{ik}$. The proof is based on substituting (0.4) into Lagrange equations

$$(0.6) \quad \frac{d}{dt}\frac{\partial L(x,\dot x)}{\partial \dot x_i} = \frac{\partial L(x,\dot x)}{\partial x_i}, \quad i = 1, \ldots, N$$

and showing that the resulting equations have the Hamitonian form (0.3).

The aim of this note is to generalize the expression (0.4) to the general case of equations (0.1) for which, as we know, the representation (0.2) is not always possible. What should be the analog of the Hamilton function $H(x)$ in this case? What should be the analog of the momenta $P_i(x)$? These are the main questions we are going to address in this note.

## 0.4. The structure of the paper

The paper is organized as follows. In Section 1 we will examine the integrals of motion associated with the dynamical equations of the most generic form (0.1). In Section 2, we will use these integrals of motion as building blocks for constructing the Lagrange function, and then, substituting it into the Lagrange equations, will show their equivalence to the equations of motion we started with. In Section 3 we construct the analogs of the Hamilton equations and discuss the possibility of using them for introducing the analogs of Poisson brackets, which, in turn would allow us to reproduce some features of the classical Hamilton formalism. After that, in Section 4, we discuss the action functional, and a little bit untypical role played by the stationary action principle in deriving the Lagrange equations. We also discuss the meaning of the Hamilton-Jacobi-type equations arising in this connection. Section 5 is completely devoted to the general coordinate transforms and the problem of finding the best coordinate system. We also consider the evolution-driven coordinate transforms and discuss in this connection the generalization of the Liouville's theorem. In Section 6 we consider a few concrete examples of explicitly building the Lagrange functions for damped harmonic oscillators in different dimensions. In conclusion we will discuss some open problems related to the uniqueness of the resulting Lagrangian. The short summary of the results is given in the last Section 7.

# 1. Invariants
## 1.1. The general solution

The general solution of system (0.1) can be written symbolically in the form

$$(1.1) \quad x_i = \exp\left(t\sum_{n=1}^{N} R_n(a)\frac{\partial}{\partial a_n}\right) a_i, \quad i = 1, \ldots, N$$

i.e., as the evolution operator acting on the vector of initial conditions $a_i = x_i(0)$, $i = 1, \ldots, N$.

From (1.1) it follows that the most general solution of system (0.1) is parametrized by $N$ arbitrary constants $c_1, \ldots, c_N$ dependent on (and fully determined by) $N$ independent initial conditions $a_1, \ldots, a_N$. From the stationarity of (0.1) it follows that shifting in time any solution of (0.1) should give another solution of the same system (0.1). This means that one of these parameters (let us select $c_N$ for that role) should play the role of the reference point for time $t$, so, it must necessarily enter in the expression for the general solution in the combination with time, as:

(1.2) $\quad x_i = X_i(c_1, \ldots, c_{N-1}, c_N + t), \quad i = 1, \ldots, N$

We are not going to discuss here the explicit constructability of all the above functions, which, in most cases, cannot be done analytically. The only thing that we need is their general structure – the way they are parametrized by $N$ arbitrary constants $c_1, \ldots, c_N$.

## 1.2. The integrals of motion

Let us treat the above system as a system of $N$ ordinary (not differential!) equations for $N$ unknowns $c_1, \ldots, c_{N-1}$ and $c_N + t$. After solving it we can (formally) express these unknowns via the coordinates $x_1, \ldots, x_N$ staying in the left-hand side of (1.2). This will lead us to the following $N$ relations:

(1.3) $\quad c_a = C_a(x), \quad a = 1, \ldots, N-1; \quad c_N + t = C_N(x)$,

in which the first $N-1$ functions $C_a(x)$, $a = 1, \ldots, N-1$ will, by construction, represent the independent conservation laws associated with a given dynamical system. As system evolves, they will not change over time, so their total time-derivatives along the trajectories of the system will vanish. The role of the last function $C_N(x)$ is different. As follows form its definition, it will change over time, but colinearly with time $t$, so its total time-derivative will be just one. We can summarize all the above statements in a single formula:

(1.4) $\quad \dot{C}_a(x) = \delta_{aN}, \quad a = 1, \ldots, N$

Unfolding (1.4) by the chain rule and using the equations (0.1) we arrive at the identities:

(1.5) $\quad \sum_{k=1}^{N} \frac{\partial C_a(x)}{\partial x_k} R_k(x) = \delta_{aN}, \quad a = 1, \ldots, N$

which can also be treated as a system of partial differential equations for functions $C_a(x)$ in which the defining functions $R_i(x)$ are assumed to be given. We are not going to discuss here the explicit constructability of all the above functions, which, again, in most cases, cannot be done analytically. However, for sufficiently smooth functions $R_i(x)$ their existence can be proved mathematically, and these functions can always be constructed approximately with any desired level of accuracy.

A few words about terminology. The $N-1$ functions $C_1(x), \ldots, C_{N-1}(x)$ (on one side), and a single function $C_N(x)$ (on another side) will play equally important but conceptually different roles in our further constructions. To stress the difference between these two groups it would be good to use for them different notations and naming conventions. Hereafter, all standard integrals of motion – i.e., functions of coordinates not changing over time on the trajectories of the system we will denote by $H_a(x)$ and call the 'space-like integrals of motion' (or 'H-integrals'), while the functions of coordinates changing on the trajectories of the system as time, we will denote by $T_a(x)$ and call them the 'time-like integrals of motion' (or 'T-integrals').

Any combination of functions $C_1(x), \ldots, C_{N-1}(x)$ is a space-like integral. This may give us $N-1$ functionally independent H-integrals:

(1.6) $\qquad H_a(x) = h_a(C_1(x), \ldots, C_{N-1}(x)), \quad a = 1, \ldots, N-1$

Also, any combination of functions $C_1(x), \ldots, C_{N-1}(x)$ added to $C_N(x)$ gives a certain time-like integral. And, as before, we may have $N-1$ functionally independent T-integrals:

(1.7) $\qquad T_a(x) = C_N(x) + g_a(C_1(x), \ldots, C_{N-1}(x)), \quad a = 1, \ldots, N-1$

For H-integrals, the following equations hold by definition:

(1.8) $\qquad \sum_{k=1}^{N} \frac{\partial H_a(x)}{\partial x_k} R_k(x) = 0, \quad a = 1, \ldots, N-1$

Similarly, for T-integrals we have:

(1.9) $\qquad \sum_{k=1}^{N} \frac{\partial T_a(x)}{\partial x_k} R_k(x) = 1, \quad a = 1, \ldots, N-1$

The $2N-2$ functions $H_a(x)$ and $T_a(x)$ are the main building blocks we will need for writing down the explicit form of the desired Lagrange function $L(x, \dot{x})$.

### 1.3. The 'field tensor'

The space- and time-like integrals of motion can also be used for defining another important set of objects represented by a collection of $N-1$ anti-symmetric tensors:

(1.10) $\qquad F_{a,ik}(x) = \frac{\partial H_a(x)}{\partial x_i}\frac{\partial T_a(x)}{\partial x_k} - \frac{\partial T_a(x)}{\partial x_i}\frac{\partial H_a(x)}{\partial x_k}, \quad a = 1, \ldots, N-1$

Multiplying both sides of these tensors by vector $R_k(x)$, summing over $k$ and using (1.8) and (1.9), we obtain:

(1.11) $\qquad \sum_{k=1}^{N} F_{a,ik}(x) R_k(x) = \frac{\partial H_a(x)}{\partial x_i}, \quad a = 1, \ldots, N-1$

Functions $F_{a,ik}(x)$ will help us to derive the corresponding equations of motion and prove that their form coincides with the form of dynamical equations (0.1) we started with. In what follows, the most important role will be played by the sum of tensors (1.10):

$$(1.12) \quad F_{ik}(x) = \sum_{a=1}^{N-1} F_{a,ik}(x) = \sum_{a=1}^{N-1} \left( \frac{\partial H_a(x)}{\partial x_i} \frac{\partial T_a(x)}{\partial x_k} - \frac{\partial T_a(x)}{\partial x_i} \frac{\partial H_a(x)}{\partial x_k} \right)$$

which we will call the 'field-tensor' for the reasons which will be explained below.

## 2. Lagrangian
### 2.1. Constructing Lagrangian

Now we have everything we need for writing down the Lagrange function for system (0.1). Starting with the $N-1$ functionally independent pairs of H- and T-integrals we introduced earlier, let us first introduce the following $N-1$ functions of coordinates and velocities

$$(2.1) \quad L_a(\dot{x}, x) = H_a(x) \left( \sum_{k=1}^{N} \frac{\partial T_a(x)}{\partial x_k} \dot{x}_k - 1 \right), \quad a = 1, \ldots, N-1$$

and then construct their sum:

$$(2.2) \quad L(x, \dot{x}) = \sum_{a=1}^{N-1} L_a(\dot{x}, x)$$

Below we will show that under some sufficiently mild restrictions, this sum represents the desired Lagrange function for system (0.1), or, more exactly – a multi-parametric family of such functions.

### 2.2. Deriving equations of motion

Let us now derive the restrictions under which the Lagrange equations of motion associated with Lagrangian (2.2) coincide with the original equations (0.1). With (2.1) and (2.2) the Lagrange equations can be written as:

$$(2.3) \quad \frac{d}{dt} \frac{\partial L(x, \dot{x})}{\partial \dot{x}_i} - \frac{\partial L(x, \dot{x})}{\partial x_i} = \sum_{a=1}^{N-1} \left( \frac{d}{dt} \frac{\partial L_a(x, \dot{x})}{\partial \dot{x}_i} - \frac{\partial L_a(x, \dot{x})}{\partial x_i} \right) = 0$$

Let us first calculate the $N-1$ elements of the above sum and then sum them up. We have:

$$(2.4) \quad \frac{\partial L_a(x, \dot{x})}{\partial x_i} = \frac{\partial H_a(x)}{\partial x_i} \sum_{k=1}^{N} \frac{\partial T_a(x)}{\partial x_k} \dot{x}_k + H_a(x) \sum_{k=1}^{N} \frac{\partial^2 T_a(x)}{\partial x_i \partial x_k} \dot{x}_k - \frac{\partial H_a(x)}{\partial x_i}$$

$$(2.5) \quad \frac{\partial L_a(x, \dot{x})}{\partial \dot{x}_i} = H_a(x) \frac{\partial T_a(x)}{\partial x_i}$$

Taking total time derivative of (2.5):

(2.6) $$\frac{d}{dt}\frac{\partial L_a(x,\dot{x})}{\partial \dot{x}_i} = H_a(x)\sum_{k=1}^{N}\frac{\partial^2 T_a(x)}{\partial x_i \partial x_k}\dot{x}_k + \frac{\partial T_a(x)}{\partial x_i}\sum_{k=1}^{N}\frac{\partial H_a(x)}{\partial x_k}\dot{x}_k,$$

combining (2.4) and (2.6) and using the definition (1.10) and formulas (1.11), we obtain:

(2.7) $$\frac{d}{dt}\frac{\partial L_a(x,\dot{x})}{\partial \dot{x}_i} - \frac{\partial L_a(x,\dot{x})}{\partial x_i} = -\sum_{k=1}^{N} F_{a,ik}(x)(\dot{x}_k - R_k(x))$$

Now, substituting (2.7) into the right-hand side of (2.3), summing up over $a$ and using the definition (1.12), we get:

(2.8) $$\frac{d}{dt}\frac{\partial L(x,\dot{x})}{\partial \dot{x}_i} - \frac{\partial L(x,\dot{x})}{\partial x_i} = -\sum_{k=1}^{N} F_{ik}(x)(\dot{x}_k - R_k(x)) = 0$$

Finally, let us look at the right-hand side of formula (2.8) containing the anti-symmetric field-tensor $F_{ik}(x)$. Remember that the determinant of any odd-dimensional anti-symmetric matrix is zero. This means that if the dimension of our system $N$ is odd, the anti-symmetric matrix $F_{ik}(x)$ cannot be inverted. However, if the dimension $N$ is even, the matrix $F_{ik}(x)$ can always be made invertible by appropriately choosing functions $h$ and $g$ in formulas (1.6) and (1.7). We will demonstrate this later, in Section 5. Now, we simply take this fact for granted and assume that the dimension of system (0.1) is even, $N = 2M$, and the condition of invertibility is satisfied. But if so, then we can assert that the system (2.8) is equivalent to the system of equations (0.1) we started with:

(2.9) $$\dot{x}_i = R_i(x), \quad i = 1, \dots, 2M$$

This completes the derivation.

### 2.3. The main theorem

Summarizing everything we can formulate the following main statement. The Lagrange function for the system (2.9) exists in even dimensions $N = 2M$ and has the form:

(2.10) $$L(x,\dot{x}) = \sum_{k=1}^{2M} P_k(x)\dot{x}_k - H(x)$$

where

(2.11) $$P_k(x) = \sum_{a=1}^{2M-1} H_a(x)\frac{\partial T_a(x)}{\partial x_k}$$

and

(2.12) $$H(x) = \sum_{k=1}^{2M-1} H_a(x)$$

Here $H_a(x), T_a(x)$, $1,\ldots,2M-1$ are any sets of the independent space- and time-like integrals of motion of system (0.1)/(2.9) ensuring the non-degeneracy of 'field tensor' (1.12).

How many different Lagrangians of type (2.10) may exist for a given system of dynamical equations (0.1)? The answer is obvious: their number is infinite, while the 'cardinality' of this infinity is determined by the number of different $H_a(x), T_a(x)$ pairs with invertible $F_{ik}(x)$. The latter number is, in turn, determined by the number of independent functions $g_a$ and $h_a$ in definitions (1.6) and (1.7) of these pairs.

### 2.4. Lagrangian on true trajectories

Assume that $x$ satisfies the equations of motion and so lies on a true trajectory of the system. What is the value of the Lagrangian on this trajectory? By using the representation

$$(2.13) \quad L(x, \dot{x}) = \sum_{a=1}^{N-1} H_a(x) \left( \sum_{k=1}^{N} \frac{\partial T_a(x)}{\partial x_k} \dot{x}_k - 1 \right) = \sum_{a=1}^{N-1} H_a(x)(\dot{T}_a(x) - 1)$$

directly following from formulas (2.1) and (2.2), one can see that the Lagrangian vanishes on the evolution trajectory:

$$(2.14) \quad L(x, \dot{x}) = 0$$

This is a trivial consequence of the fact that on any evolution trajectory $\dot{T}_a(x) = 1$ for $a = 1, \ldots, N-1$.

### 2.5. The odd-dimensional case

Although, as we showed above, the direct construction of the Lagrange function for the odd-dimensional equations (0.1) is impossible, this function can be constructed indirectly, after extending the dimension of the original equations by 1. As we show below, this extension can always be done in the 'minimalist' ways not affecting the form of the original (odd-dimensional) system of equations and its solutions.

The idea of such an extension is very simple: consider a $(2M-1)$-dimensional system of equations (0.1) with the right-hand sides additionally depending on a (constant) numeric parameter $a$:

$$(2.15) \quad \dot{x}_i = R_i(x, a), \quad i = 1, \ldots, 2M-1$$

Let's treat this numerical parameter as a constant dynamical variable $x_{N+1} = a$. Then the extended $2M$-dimensional system will have the form

$$(2.16) \quad \dot{x}_i = R_i(x_1, \ldots, x_{2M}), \quad i = 1, \ldots, 2M-1, \quad \dot{x}_{2M} = 0$$

Note that no explicit parameterization of right-hand sides of equations (0.1) is required. The actual dependency on the numeric parameter $a$ does not have to be present in the original system. It could be introduced artificially as a trivial extension like:

$$(2.17) \quad R_i(x, x_{2M}) = R_i(x) + x_{2M}, \quad i = 1, \ldots, 2M-1, \quad x_{2M} = 0$$

or

(2.18)  $R_i(x, x_{2M}) = R_i(x)x_{2M}, \quad i = 1, \ldots, 2M-1, , \quad x_{2M} = 1$

Such extensions obviously do not change the form of equations (0.1) and/or their solutions, but allow one to increase the dimension of the system by one making it odd-dimensional in the mildest possible way.

As a simplest illustrative example, consider the 1-dimensional equation

(2.19)  $\dot{x} = R(x).$

Although we can easily construct the time-like integral of motion $T = \int dx R^{-1}(x)$ for it, we cannot build the Lagrangian because of the absence of a non-trivial space-like integrals of motion $H(x)$ (we simply cannot have it as our system is 1-dimensional). But we can improve the situation by transforming the constant parameters of this equation into an additional variable. Although we don't see any additional parameters in function $R(x)$, we can easily introduce it in many ways. Consider, for example, the following extension of function $R(x)$: $R(x) \to 1 \cdot R(x) \to yR(x)$, which, obviously, corresponds to the case (2.18). Then our equation transforms into the system: $\dot{x} = yR(x), \quad \dot{y} = 0$. Now the time-like integral of motion will take the form: $T = y^{-1}\int dx R^{-1}(x)$. As the spacelike integral of motion $H(x)$ we can take any function of $y$, say $H = y^2$. In this case, the Lagrangian is well-defined and gives

(2.20)  $L = \dfrac{y\dot{x}}{R(x)} - \dot{y}\int \dfrac{dx}{R(x)} - y^2$

It is easy to see that applying to it the Lagrange equations we obtain the system (2.19).

## 3. Hamiltonian
### 3.1. The 'Hamiltonian' and 'momenta'

The functions $P_i(x)$ in the expression (2.11) look like classical momenta. They are defined by the same formula as in the classical case:

(3.1)  $P_k(x) = \dfrac{\partial L(x, \dot{x})}{\partial \dot{x}_k}, \quad k = 1, \ldots, 2M$

As to $H(x)$ – it looks like the classical Hamiltonian. First of all, as follows from (2.12), it is a conserved quantity. Second, it is related to the Lagrangian (we just constructed) exactly in the same ways as in the classical case. Indeed, by using explicit formula (2.10) for $L(x, \dot{x})$ we just derived, we can show that

(3.2)  $H(x) = \sum_{k=1}^{2M} \dot{x}_k \dfrac{\partial L(x, \dot{x})}{\partial \dot{x}_k} - L(x, \dot{x})$

However, $H(x)$ is, obviously, not a classical Hamiltonian in case of non-conservative systems.

## 3.2. The 'Hamilton equations'

To see what the difference is, let us sum up both sides of (1.11) by $a = 1, \ldots, 2M - 1$. We obtain the equation

$$(3.3) \qquad \sum_{k=1}^{2M} F_{ik}(x) R_k(x) = \frac{\partial H(x)}{\partial x_i}, \quad i = 1, \ldots, 2M$$

Defining the inverse of matrix (1.12) as

$$(3.4) \qquad \sigma_{ik}(x) = F_{ik}^{-1}(x)$$

we can rewrite (3.3) in the form:

$$(3.5) \qquad \dot{x}_i = \sum_{k=1}^{2M} \sigma_{ik}(x) \frac{\partial H(x)}{\partial x_k}, \quad i = 1, \ldots, 2M$$

This form resembles the form of the Hamilton equation (0.3) for conservative systems. The only difference is that the anti-symmetric tensor $\sigma_{ik}(x)$ is now a function of dynamical variables $x$.

## 3.3. The 'Poisson brackets'

The equation (3.5) allows one to calculate the total time-derivative of any function of dynamical variables. Let $f(x)$ be such a function. Then differentiating it and using (3.5) we obtain:

$$(3.6) \qquad \dot{f}(x) = \sum_{k=1}^{2M} \frac{\partial f(x)}{\partial x_i} \dot{x}_i = \sum_{k=1}^{N} \sigma_{ik}(x) \frac{\partial f(x)}{\partial x_i} \frac{\partial H(x)}{\partial x_k}$$

We can define the following binary operation:

$$(3.7) \qquad \{f(x), g(x)\} = \sum_{i,k=1}^{2M} \sigma_{ik}(x) \frac{\partial f(x)}{\partial x_i} \frac{\partial g(x)}{\partial x_k}$$

resembling the form of the classical Poisson bracket. Using this bracket we can rewrite (3.6) as:

$$(3.8) \qquad \dot{f}(x) = \{f(x), H(x)\}$$

In terms of this operation, one can say that a certain function $f(x)$ is conserved if its Poisson bracket with Hamiltonian is zero. This again makes $H(x)$ like the classical Hamiltonian and gives another way of seeing that the latter is a conserved quantity. As direct consequence of (3.8) we can write down the following relations:

$$(3.9) \qquad \{H_a(x), H(x)\} = 0, \quad \{T_a(x), H(x)\} = 1$$

Also, the commutation relations for the coordinates $x_i$ have the form:

(3.10) $$\{x_i, x_k\} = \sigma_{ik}(x)$$

which resemble the Poisson relations for the Hamiltonian case, with the only exception that now their right-hand sides are not constants but some non-trivial functions of $x$. Note that (3.10) allows one to rewrite the relation (3.7) in the form

(3.11) $$\{f(x), g(x)\} = \sum_{i,k=1}^{2M} \{x_i, x_k\} \frac{\partial f(x)}{\partial x_i} \frac{\partial g(x)}{\partial x_k}$$

which exactly coincides with the corresponding classical relation. It is easy to see that the brackets (3.7) have many properties of the standard Poisson brackets. First, they are anti-symmetric:

(3.12) $$\{f(x), g(x)\} = -\{g(x), f(x)\}$$

Second, they obey the multiplication property:

(3.13) $$\{f(x), g(x)h(x)\} = g(x)\{f(x), f(x)\} + h(x)\{f(x), g(x)\}$$

Finally, the satisfy the Jacobi identity:

(3.14) $$\{f(x), \{g(x), h(x)\}\} + \{g(x), \{h(x), f(x)\}\} + \{h(x), \{f(x), g(x)\}\} = 0$$

The proof of the latter relation will be given below, in the next section.

### 3.4. The 'Jacobi identity'

To prove (3.14) let us start with the relation

(3.15) $$\{f(x), \{g(x), h(x)\}\} = U(f, g, h) + V(f, g, h)$$

in which

(3.16) $$U(f, g, h) = \sum_{i,k,l,m=0}^{N} \sigma_{ik}(x) \frac{\partial \sigma_{lm}(x)}{\partial x_k} \left( \frac{\partial f(x)}{\partial x_i} \frac{\partial g(x)}{\partial x_l} \frac{\partial h(x)}{\partial x_m} \right),$$

and

(3.17) $$V(f, g, h) = \sum_{i,k,l,m=0}^{N} \sigma_{ik}(x)\sigma_{lm}(x) \left( \frac{\partial f(x)}{\partial x_i} \frac{\partial g(x)}{\partial x_l} \frac{\partial^2 h(x)}{\partial x_m \partial x_k} + \frac{\partial f(x)}{\partial x_i} \frac{\partial^2 g(x)}{\partial x_l \partial x_k} \frac{\partial h(x)}{\partial x_m} \right)$$

From the definition of tensor $\sigma_{ik}(x)$ as of inverse of $F_{ik}(x)$ it follows that

(3.18) $$\sigma_{ik}(x) \frac{\partial \sigma_{lm}(x)}{\partial x_k} = \sigma_{ik}(x)\sigma_{ls}(x)\sigma_{mr}(x) \frac{\partial F_{sr}(x)}{\partial x_k}$$

This immediately gives us

(3.19) $$U(f,g,h) = \sum_{i,k,l,m=0}^{N} \sigma_{ik}(x)\sigma_{ls}(x)\sigma_{mr}(x)\left(\frac{\partial f(x)}{\partial x_i}\frac{\partial g(x)}{\partial x_l}\frac{\partial h(x)}{\partial x_m}\right)\frac{\partial F_{sr}(x)}{\partial x_k}$$

From the definition of tensor $F_{sr}(x)$ it follows that

(3.20) $$\frac{\partial F_{sr}(x)}{\partial x_k} + \frac{\partial F_{rk}(x)}{\partial x_s} + \frac{\partial F_{ks}(x)}{\partial x_r} = 0$$

(the cyclic property). Using it we can easily obtain that

(3.21) $$U(f,g,h) + U(g,h,f) + U(h,f,g) = 0$$

It is also easy to show that

(3.22) $$V(f,g,h) + V(g,h,f) + V(h,f,g) = 0$$

Combining these two cyclic relations for $U$ and $V$, we obtain (3.14).

## 4. Action
### 4.1. The variation problem: too many initial conditions

The interpretation of Lagrange equations as a condition for an extremum of the action functional

(4.1) $$S = \int_{t_1}^{t_2} L(x,\dot{x})dt$$

leads to the variational condition:

(4.2) $$\delta S = \sum_{k=1}^{N} \delta x_i \frac{\partial L(x,\dot{x})}{\partial \dot{x}_i}\bigg|_{t_1}^{t_2} + \int_{t_1, x(t_1)=a}^{t_2, x(t_2)=b} \sum_{i=1}^{N} \delta x_i \left(\frac{d}{dt}\frac{\partial L(x,\dot{x})}{\partial \dot{x}_i} - \frac{\partial L(x,\dot{x})}{\partial x_i}\right) dt = 0$$

whose fulfillment requires not only that the Lagrange equations should be satisfied, but also that the two boundary terms in (4.2) should vanish. In our case of the Lagrangian given by (2.10) this would give us the following $2N$ conditions

(4.3) $$\sum_{i=1}^{N}(P_i(x)\delta x_i)_{t=t_1} = 0, \quad \sum_{i=1}^{N}(P_i(x)\delta x_i)_{t=t_2} = 0$$

Traditional ways of ensuring this is via the conditions

(4.4) $$\delta x_i(t_1) = \delta x_i(t_2) = 0, \quad i = 1, \dots, N$$

These conditions are essentially equivalent to imposing the $N$ start-point and $N$ end-point conditions on the solutions of dynamical equations. And this would be completely fine if we would use the more traditional Lagrange functions quadratic in velocities and thus leading to the second order equations. Indeed, the general solutions of the systems of $N$ second-order equations are parametrized by $2N$ arbitrary parameters and imposing on them the $2N$ constraints is a quite legitimate operation. However, our case is different. We have a system of first-order equations whose solutions are parametrized by only $N$ parameters and imposing on them $2N$ constraints would be too much. What can we do? The answer to this question is based on the existence of many equivalent Lagrange functions associated with the same set of dynamical equations.

### 4.2. The 'gauge invariance'

The equations of motion are invariant under all transforms preserving the form of the anti-symmetric tensor $\sigma_{ik}(x)$ – the inverse of the tensor $F_{ik}(x)$ defined in (2.9). To see how these transforms may look like, let's apply the Lagrange equations to the Lagrange function (2.11). Comparing the result with (3.3), we find another equivalent expression of tensor $F_{ik}(x)$ directly via the momenta $P_i(x)$:

(4.5) $\qquad F_{ik}(x) = \dfrac{\partial P_i(x)}{\partial x_k} - \dfrac{\partial P_k(x)}{\partial x_i}$

This expression resembles the form of the field tensor in classical electrodynamics and similarly to the latter will not change after the replacement

(4.6) $\qquad P_i(x) \to \bar{P}_i(x) = P_i(x) + \dfrac{\partial \Phi(x)}{\partial x_i}$

We will call (4.5) the 'gauge transform' for its similarity with the analogous transform in classical electrodynamics. In this sense, the functions $P_i(x)$ are the analogs of the vector-potential. The form of the corresponding equations of motion are obviously invariant under transform (4.5).

From the identity:

(4.7) $\qquad \dfrac{d\Phi(x)}{dt} = \sum_{k=1}^{N} \dfrac{\partial \Phi(x)}{\partial x_k} \dot{x}_k$

it follows that performing the gauge transform is equivalent to adding the total time-derivative to the Lagrangian:

(4.8) $\qquad L(x,\dot{x}) \to \bar{L}(x,\dot{x}) = L(x,\dot{x}) + \dfrac{d\Phi(x)}{dt}$

In the traditional (i.e., the second-order) Lagrange formalism, the function $\Phi(x,t)$ can be chosen arbitrarily. Our case is different. In our case the form of the function $\Phi(x,t)$ is dictated by the necessity of 'saving' the derivability of Lagrange equations from the principle of extremal action.

## 4.3. 'Hamilton-Jacobi equation' as 'gauge fixing' condition

The gauge transform plays an especially important role in the first-order Lagrange formalism. It enters the game as a rescuer solving the problem with too many initial conditions. Indeed, note that the nullification of the boundary terms can be achieved not only via nullifying the variations (which is equivalent to imposing constraints on solutions) but also via nullifying the terms $\partial L/\partial \dot{x}_i$, which in our case are equal to the vector-potentials $P_i(x)$. But these terms are not gauge invariant. Therefore, we always can find such a function $\Phi(x)$ which would ensure the validity of $N$ equations

$$(4.9) \qquad \bar{P}_i(x_2) = \left(P_i(x) + \frac{\partial \Phi(x)}{\partial x_i}\right)_{x=x_2, t=t_2} = 0$$

for the gauge-transformed vector-potentials at one specific instance of time $t = t_2$ and corresponding coordinates $x = x_2$. This is essentially the analog of the gauge-fixing condition in classical electrodynamics. Condition like (4.10) cannot be imposed on both ends of the trajectory, therefore at the lower end of integration interval we may still have the standard initial conditions

$$(4.10) \qquad x_i(t_1) = a_i, \quad i = 1, \ldots, N$$

The set of legitimate functions $\Phi(x)$ can be described as follows. First, we need to express the Hamiltonian $H(x)$ via the 'momenta' variables

$$(4.11) \qquad p_i = P_i(x)$$

defining the new function $\mathcal{H}(p)$ as

$$(4.12) \qquad \mathcal{H}(p) = \mathcal{H}(P(x)) = H(x)$$

and then replace the variables $p_i$ with partial derivatives $\partial \Phi(x)/\partial x_i$. This will give us the following first-order partial differential equation for function $\Phi(x)$

$$(4.13) \qquad \mathcal{H}\left(\frac{\partial \Phi(x)}{\partial x_1}, \ldots, \frac{\partial \Phi(x)}{\partial x_N}\right) = const$$

resembling the form of the classical Hamilton-Jacobi equation for the stationary case.

The simplest explicit form of the gauge-fixing function (which can easily be found from the above equation) is given by:

$$(4.14) \qquad \Phi(x) = \sum_{i=1}^{N} P_i(x_2) x_i$$

The corresponding 'corrected' – i.e. fully legitimate form of the action-principle compatible Lagrangian is

$$(4.15) \qquad \bar{L}(x, \dot{x}) = \sum_{k=1}^{N} (P_k(x) - P_k(x_2)) \dot{x}_k - H(x)$$

It is easily seen that the upper-limit problem for this lagrangian is automatically solved.

## 4.4. Action as function of upper-limit parameters

If we take the integral of the extended (corrected) Lagrangian (4.15) and note that the on the trajectory of the system and noting that the non-extended (non-corrected) Lagrangian on the trajectory vanishes, we obtain that

(4.16) $$\int_{t_1}^{t_2} \bar{L}(x, \dot{x}) dt = -\Phi(x_2) + \Phi(x_1)$$

This means that the gauge-fixing function $\Phi(x)$ has the meaning of the action integral (with sign minus) considered as a function of upper-end integration parameters provided that the system trajectory is optimal. This is a direct consequence of the null-property of the Lagrangian (2.14) we discussed in section 2.

Note that without such extension (i.e., with the uncorrected Lagrangian), the existence of the analogies with the classical Hamilton-Jacobi formalism would be impossible, because, if on the trajectories of the system the Lagrangian is identically zero, then the action functional should also be identically zero on the trajectories of the system and thus cannot be non-trivial function of coordinates and time. However, the gauge invariance allowed us to correctly address and solve this problem.

# 5. Coordinates
## 5.1. The 'best' coordinate system

All mathematical constructions described above (including both the original equations of motion as well as the corresponding Lagrangians and Lagrange/Hamilton-type equations) preserve their form after transforming the original coordinates $x = \{x_1, \ldots, x_N\}$ into the new coordinates $y = \{y_1, \ldots, y_N\}$:

(5.1) $\quad y = f(x)$

The transition to the new coordinates may change the form of the functions $R_i(x)$ defining the right-hand-sides of dynamical equations, and all the corresponding functions, which include: the form of momenta $P_i(x)$ and Hamiltonian $H(x)$ determining the form of Lagrangian, the $H_a(x)$ and $T_a(x)$ and many others. In some coordinates these functions may look more complex, in others – simpler. The natural question arising in this connection is if there exists a distinguished system of coordinates in which the dynamical equations and their derivatives have a maximally simple form – i.e., a form not allowing any further simplification. The answer to this question has already been given in section 1: the equations (1.13) describing the transition to space- and time-like integrals of motion describe such a coordinate transform. The left-hand sides of (1.13) can be treated as the new dynamical variables:

(5.2) $\quad y_i = C_i(x), \quad i = 1, \ldots, N$

The equations of motion in these variables have the form:

(5.3) $\quad \dot{y}_i = R_i(y)$

with

(5.4) $\quad R_i(y) = \delta_{iN}, \quad i = 1, \ldots, N$

This is the simplest possible system of coordinates in which any dynamical system can be rewritten after applying to it an appropriate coordinate transformation. We call it the 'best system of coordinates' and denote by $D$. From this representation it follows that in its very core each dynamical system is trivial because it is equivalent to distinguished system: one of whose variables is simply time, and others are constants. Also, from this standpoint any two dynamical systems having the same dimension are equivalent: indeed, if any of these two systems $S_1$ and $S_2$ can be reduced to the distinguished form (5.3)/(5.4) with an invertible coordinate transforms $U_{1D}$ and $U_{2D}$, this means that there must be a transform $U_{12} = U_2 U_1^{-1}$ converting system $S_1$ into $S_2$.

## 5.2. Lagrangian in the 'best' coordinate system

The fact that for any system of equations of motion it is possible to find the 'best' coordinate system in which the equations of motion take the simplest possible form (5.3)/(5.4) is remarkable because the construction of the Lagrange function for such a form becomes trivial. The corresponding Lagrangian in this case can be constructed by means of the general prescriptions given in Section 2. Define

(5.5) $\quad H_i = y_i, \quad T_i = y_N + y_{i+1}, \quad i = 1, \ldots, N-2; \quad H_{N-1} = y_{N-1}, \quad T_{N-1} = y_N + y_1$

Then we will have

(5.6) $\quad P_1(y) = y_{N-1}, \quad P_i(y) = y_{i-1}, \quad i = 2, \ldots, N-1, \quad P_N(y) = H(y) = \sum_{i=1}^{N-1} y_i,$

which completely defines the Lagrangian in the coordinates $y$. This also means that after performing the 'back-transform' to the original coordinates $x$, we obtain the Lagrange function for the original system of equations in the original coordinates. This gives us the proof of the existence of Lagrange function for any even-dimensional system of first-order dynamical equations.

One interesting observation in this connection. We can immediately see that in the best coordinate system and for the even values of $N$ formulas (5.6) define a Hamilton-type system. Indeed, the anti-symmetric field tensor $F_{ik}$ is in this case invertible, and the fact that it is a constant tensor

(5.7) $\quad F_{i,i-1} = -F_{i-1,i} = 1, \quad F_{1,N-1} = -F_{N-1,1} = 1, \quad F_{N,i} = -F_{i,N} = 1$

means that its inverse, $\sigma_{ik}$, is also a constant anti-symmetric tensor. Thus, according to (0.2), this is a typical Hamilton-type system. This fact shouldn't surprise us because the general coordinate transform is not expected to preserve the Hamilton character of a system. It also does not preserve the 'conservativity' of those systems for which the notion of 'energy' can be introduced in meaningful ways. This is because the scope of the general coordinate transform is much broader than that of the canonical transform preserving the conservativity (and Hamiltonian property) of a system by construction.

## 5.3. How to construct the 'best' coordinate system?

Let us consider the system of coordinates $x(t) = \{x_1(t), \ldots, x_N(t)\}$ as functions of time. Assume that we want to transform this system to the 'best' form $\xi(t) = \{\xi(t), \ldots, \xi_N(t)\}$, i.e., to a system of variables with trivial dependence on time: $\xi_i(t) = c_i$, $i = 1, \ldots, N-1$ and $\xi_N(t) = t + c_N$. Let

(5.8) $\qquad \xi_i(t) = \Xi_i(x(t)), \quad i = 1, \ldots, N$

represent such a transform parametrized by some unknown functions $\Xi_i(x)$, $i - 1, \ldots, N$. Our goal is to determine their form, provided that the form of functions $x_i(t)$, $i = 1, \ldots, N$ is given. We can think of the latter as of functions obtained during direct observations on dynamical variables $x$ at different instances of time.

We want components $\Xi_1, \ldots, \Xi_{N-1}$ to be constants and $\Xi_0$ behaving as $t$. We can rephrase this as a requirement of minimizing the following expression over some time interval $[t_1, t_2]$

(5.9) $\qquad S = \int_{t_1}^{t_2} \left\{ \left(\frac{d\Xi_N}{dt} - 1\right)^2 + \sum_{i=1}^{N-1} \left(\frac{d\Xi_i}{dt}\right)^2 \right\} dt$

This expression can be simplified to the sum of the term

(5.10) $\qquad S = \int_{t_1}^{t_2} \sum_{i=0}^{N} \left(\frac{d\Xi_i}{dt}\right)^2 dt$

and some inessential additional constant terms

(5.11) $\qquad \Delta S = -2(\xi_0(t_2) - \xi_0(t_1)) + (t_2 - t_1) = -(t_2 - t_1)$

arising as a result of integration of total time derivatives and usage of boundary conditions

(5.12) $\qquad \xi_i(t_A) = \delta_{0A} t_A + c_i, \quad A = 1,2$

where $c_i$ are constants.

In order to solve this problem, we need to rewrite everything in terms of the original variables $x$. In the original coordinates this expression transforms into the

(5.13) $\qquad S = \int_{t_1}^{t_2} \left( g^{nm}(x) \frac{dx_n}{dt} \frac{dx_m}{dt} \right) dt$

resembling the form of the action functional for a particle on a Riemannian manifold characterized by the metric tensor

(5.14) $\qquad g^{nm}(x) = \sum_{i=0}^{N} \left(\frac{d\Xi_i(x)}{dx_n}\right)\left(\frac{d\Xi_i(x)}{dx_m}\right)$

Now, our goal is to find the family of transforms $\Xi_i(x)$ for which the value of (5.13) taken along the path in the space of observed values of $x$ would be minimal.

This a typical variational problem which can be solved approximately with any desired degree of accuracy after parametrizing the sought functions $\Xi_i$ by some sets of numerical parameters:

(5.15) $\qquad \Xi_i(x) = \Xi_i(x, a), \quad i = 1, \ldots, N$

and minimizing the functional (5.14) (which now becomes a function of unknown parameters $\alpha$) with the additional constraints

(5.16) $\qquad \Xi_i(x(t_A), a) = \delta_{0A} t_A + c_i, \quad A = 1,2$

with respect to $\alpha$.

## 5.4. The 'Liouville theorem'

In Hamilton case, as system evolves, the elementary volume of the phase space remains unchanged. This is stated by the famous Liouville's theorem following from the fact that the Jacobian of the evolution-driven coordinate transform for Hamilton-like right-hand sides $R_i(x)$ is 1. In case of general systems (0.1), the situation could be different and the conservation of the volume $dx_1 \ldots dx_N$ cannot be guaranteed. What is the analog of the Liouville's theorem is that case?

To answer this question let us compute the Jacobian of the evolution-driven transform (1.1) in the most general case. Selecting two arbitrary times $t_a$, $t_b$ and the reference time $t = 0$, we can write:

(5.17a) $\qquad x_i = f_i(t_a, x_{a1}, \ldots, x_{aN}) = \exp\left(-t_a \sum_{k=1}^{N} R_k(x_{ai}) \frac{\partial}{\partial x_{ak}}\right) x_{ai}, \quad i = 1, \ldots, N$

(5.17b) $\qquad x_i = f_i(t_b, x_{b1}, \ldots, x_{bN}) = \exp\left(-t_b \sum_{k=1}^{N} R_k(x_{bi}) \frac{\partial}{\partial x_{bk}}\right) x_{bi}, \quad i = 1, \ldots, N$

where $x_{ai} = x_i(t_a)$, $x_{bi} = x_i(t_b)$ and $x_i = x_i(0)$. Note that in both formulas the transform is represented by one and the same function $f$ of time and coordinates. For that reason, the elementary volumes in the spaces of coordinates $x_a$ and $x_b$ are related as

(5.18) $\qquad J(t_a, x_a) dx_{a1} \ldots dx_{aN} = J(t_b, x_b) dx_{b1} \ldots dx_{bN}$

where

(5.19) $\qquad J_f(t, x) = \det\left(\frac{\partial f_i(t, x)}{\partial x_k}\right)$

denotes the corresponding Jacobian. This means that the elementary volume-based object

(5.20) $$dV = J_f(t,x)dx_1 \ldots dx_N$$

is invariant under all evolution-driven coordinate transforms. Note that the scaling factor in the above expression (the Jacobian) is considered as a function of both coordinates and time. If we remember that, according to (1.3), the time variable can also be expressed via the coordinates as $t = C_N(x)$, we arrive at the specific form of formula (5.20):

(5.21) $$dV = J_f(C_N(x), x)dx_1 \ldots dx_N = J(x)dx_1 \ldots dx_N$$

in which the scaling factor $J(x)$ is now a function of coordinates only. This formula can be considered as the desired extension of the Liouville's theorem to the case of arbitrary dynamical systems.

Finally note that the formula (5.21) can also be interpreted as the information conservation law for the generic systems of the type (9.1). Of course, one needs to be cautious when quantifying the informational content of the continuous systems, because, formally, each separate point in a continuous coordinate space needs infinite memory for its exact identification. In such a situation, it is impossible to directly quantify the amount of information contained in the elementary volume of that space. It is intuitively clear, however, that the amount of information contained in such a volume should somehow be related to its size, $dx_1 \ldots dx_N$, maybe, taken with some additional scaling factor. And the factor must be chosen from the condition of the invariance of the whole expression under the evolution-driven transform. Why? Because, as it follows from the general definition of the latter transform (1.1), it is invertible which means that it cannot lead to any accumulation or loss of the information contained in dynamical variables. And he only factor guaranteeing such a strict conservation of information is the factor $J(x)$ in formula (5.21).

### 5.5. The multivalued Lagrangians

While the information is conserved during the evolution of the system, it may be not conserved in other (non-evolution-based) coordinate transforms. This may happen any time when the direct and inverse transforms are inequivalent from the standpoint of branching/merging. For example, the univalued function $y = x^2$ does not conserve the information because it maps two points with opposite sign onto one single point (merging), so its inverse $x = \pm\sqrt{y}$ is a bivalued (branching) function.

In our case, the direct map (from the Lagrangian to the equations of motion) looks univalued – we get a single system of equations for a given Lagrangian by construction. At the same time, the uniquiness of the inverse map (i.e. from the equations of motion to a Lagrangian) cannot be a-priori guaranteed in general because the transform from the original variables to the time- and space-like integrals of motion (used as components of the Lagrangian) may require solving the systems of ordinary equations – the operation which may (in some cases) give multiple results.

If that happens, we may end up with the multi-valued Lagrangians having some (maybe even infinite) number of branches which, however, must be totally equivalent in the sense that applying the Lagrange equation to each these branches should lead to the same original equations of motion. This is possible in two cases only: if the difference between the branches can be described a) by a total time-derivative of a certain function added to the Lagrangian, or b) by an appearance of a constant muliplier of the Lagrangian.

In any case, this is an interesting theoretical possibility to explore further. However, it is also clear that these multi-valued Lagrangians are the result of choosing a specific coordinate system. By simply choosing another system may reduce the Lagrangians to the normal single-valued form. For the examples of such constructions in simple models see Section 6.1.

# 6. Examples
## 6.1. One-dimensional damped harmonic oscillator

Consider the standard Newtonian equation for the stationary damped harmonic oscillator with constant frequency $\omega$ and constant damping coefficient $2\gamma$

(6.1) $\quad \ddot{x} + 2\gamma \dot{x} + \omega^2 x = 0$

There are infinitely many ways of bringing this equation to the first-order form. We will consider the most symmetric one:

(6.2a) $\quad \dot{x}_1 = -\gamma x_1 + \rho x_2$

(6.2b) $\quad \dot{x}_2 = -\gamma x_2 - \rho x_1$

in which

(6.3) $\quad \rho = \sqrt{\omega^2 - \gamma^2}$

Exclusion of each of these variables from the system leads to the equation of the form (4.1) for another one. The most general solution of this equation (which in this case can be found analytically) is given by

(6.4a) $\quad x_1 = ae^{-\gamma(t+c_2)} \cos \rho(t + c)$

(6.4b) $\quad x_2 = ae^{-\gamma(t+c_2)} \sin \rho(t + c)$

in which $a$ and $c$ are two arbitrary constants. Note that constant $c$ enters in the expressions (4.3) in combination with $t$, so it serves the reference point for time.

Considering the two above expressions as two algebraic equations for $t + c$ and $a$ and resolving them, we obtain the expressions for $T(x)$ and $H(x)$:

(6.5a) $\quad t + c = \frac{1}{\rho} \arctan\left(\frac{x_2}{x_1}\right) = T(x)$

(6.5b) $\quad a^2 = (x_1^2 + x_2^2) \exp\left\{-\frac{2\gamma}{\rho} \arctan\left(\frac{x_2}{x_1}\right)\right\} = H(x)$

Note that $H(x)$ is an integral of motion – a function of $x_1, x_2$ not changing on the trajectory of the system, while $T(x)$ is a time-like object – another function of $x_1, x_2$ behaving on the trajectory of the system like time $t$. Calculating the partial derivatives of $T(x)$:

(6.6) $$\frac{\partial T(x)}{\partial x_1} = \frac{-x_2}{x_1^2 + x_2^2}, \quad \frac{\partial T(x)}{\partial x_2} = \frac{+x_1}{x_1^2 + x_2^2}$$

and substituting them into the general expression for the Lagrange function, we obtain the explicit form of Lagrange function for damped harmonic oscillator:

(6.7) $$L(x, \dot{x}) = \frac{1}{2}(x_2 \dot{x}_1 - x_1 \dot{x}_2 - x_1^2 - x_2^2) \exp\left\{-\frac{2\gamma}{\rho} \arctan\left(\frac{x_2}{x_1}\right)\right\}$$

Note that if $\gamma \neq 0$ the system is dissipative. If $\gamma = 0$ it becomes conservative and its Lagrangian reduces to the standard Lagrangian for simple harmonic oscillator:

(6.8) $$L(x, \dot{x}) = \frac{1}{2}(x_2 \dot{x}_1 - x_1 \dot{x}_2 - x_1^2 - x_2^2)$$

Let's compare the Lagrangians for dissipative (4.8) and conservative (4.9) cases. What makes them so different? The only essential difference is that (4.8) contains the multivalued function arctan (...) having infinite number of branches. Each time when the oscillator crosses its equilibrium point at $x_1 = 0$, the switch of branches occurs and the arctan (...) function changes (increases or decreases) by a constant $\pi$. This means that all branches of the function (4.8) can be described by a constant factor $\exp(-2\pi n\gamma/\rho)$ in which $n$ is an arbitrary integer. Introduction of a constant factor does not obviously change the equations of motion, therefore all branches of this multi-branch Lagrangian lead to the same system of equations (6.2).

Note that by simple variable transform we can get rid of this multi-branch structure of the Lagrangian. Indeed, let us replace the initial variables $x_1$ and $x_2$ with their radial and angular versions $r$ and $\theta$ as

(6.9) $$x_1 = r \cos \theta, \quad x_2 = r \sin \theta$$

This change replaces the equations of motion with the much simple system:

(6.10) $$\dot{r} = -\gamma r, \quad \dot{\theta} = \rho$$

whose solutions will give us:

(6.11) $$r = a \exp\{-\gamma(t + c)\}, \quad \theta = \rho(t + c)$$

Resolving these solutions with respect to constant $a$ and the shifted time variable $t + c$, we obtain:

(6.12) $$t + c = \theta/\rho = T(r, \theta), \quad a^2 = r^2 \exp\{-2\gamma\theta\} = H(r, \theta)$$

As seen from (6.14), the partial derivatives of $T(r, \theta)$ have only one non-zero component

(6.13) $$\frac{\partial T(r, \theta)}{\partial \theta} = \frac{1}{\rho}, \quad \frac{\partial T(r, \theta)}{\partial r} = 0$$

so, their substitution into the general expression for the Lagrange function (6.7), gives

(6.14) $$L(r, \theta, \dot{r}, \dot{\theta}) = r^2 \left(\frac{\dot{\theta}}{\rho} - 1\right) \exp\{-2\gamma\theta\}$$

## 6.2. Multi-dimensional non-physical system

Consider the following system of first-order differential equations:

(6.15) $$\dot{x}_i = \sum_{k \neq i}^{N} \frac{1}{x_i - x_k}$$

In contrast with the linear models considered in previous section, this model is non-linear, does not contain any free parameters and is intrinsically non-Hamiltonian. From this equation we can recursively find all its integrals of motion expressing them via the momenta:

(6.16) $$s_n(x) = \sum_{k=1}^{N} x_k^n$$

In order to do that we start with the following obvious formula:

(6.17) $$\sum_{i,k \neq i}^{N} \frac{f(x_i)}{x_i - x_k} = \frac{1}{2} \sum_{i,k \neq i}^{N} \frac{f(x_i) - f(x_k)}{x_i - x_k}$$

Multiplying both sides of (6.15) by $F(x_i)$ taking sum over $i$ and using (6.17) we obtain:

(6.18) $$\frac{d}{dt}\left(\sum_k \int f(x_i) dx_i\right) = \frac{1}{2} \sum_{i,k \neq i}^{N} \frac{f(x_i) - f(x_k)}{x_i - x_k}$$

Now we will use the obtained relation (6.18) by taking successively

(6.19) $$f(x) = x^n, \quad n = 1, 2, \ldots$$

After some simple algebra we get the following recursion formula:

(6.20) $$\frac{ds_{n+1}(x)}{dt} = \frac{n+1}{2} \sum_{i,k \neq i}^{N} s_m(x) s_{n-m-1}(x) - \frac{n(n+1)}{2} s_{n-1}(x)$$

We see that the derivates of the higher momenta are recursively expressed via non-linear combinations of the lower momenta. Let us now examine this formula in more detail. Let us start with $n = 0$. In this case we have

(6.21) $$\frac{ds_1(x)}{dt} = 0$$

which shows us that the first momentum $s_1(x)$ is one of the integrals of motion. We use for it the standard notation:

(6.22) $\qquad C_1(x) = s_1(x)$

Taking $n = 1$ we get

(6.23) $\qquad \dfrac{ds_2(x)}{dt} = s_0^2(x) - s_0(x) = N(N-1)$

From (7.23) it follows that the second momentum is a linear function of time:

(6.24) $\qquad s_2(x) = N(N-1)t + const$

By simply rescaling it we can build the time-like integral of motion described in section 1:

(6.25) $\qquad C_N(x) = \dfrac{1}{N(N-1)} s_2(x)$

It is remarkable that the knowledge of $H_1(x)$ and $T(x)$ integrals allows one to recursively reconstruct all others. For example, taking $n = 2$ we get

(6.26) $\qquad \dfrac{ds_3(x)}{dt} = 3s_1(x)s_0(x) - 3s_1(x) = 3(N-1)s_1(x)$

which results in relation

(6.27) $\qquad s_3(x) = 3(N-1)s_1(x)t + const$

leading to the second space-like integral:

(6.28) $\qquad C_2(x) = s_3(x) - \dfrac{3}{N} s_2(x)s_1(x)$

Similarly, after taking $n = 3$, we get after some simple algebra the third space-like integral

(6.29) $\qquad C_3(x) = s_4(x) - \dfrac{2N-3}{N(N-1)} s_2^2(x) - s_1(x)s_3(x)$

This proceduree can be continued further giving us the next space-like integrals $H_n(x)$ with $n = 1, 2, \ldots, N-1$.

This defines the coordinate transform

(6.30) $\qquad y_1 = C_1(x), \ y_2 = C_2(x), \ y_3 = C_3(x), \ldots, \ y_N = C_N(x)$

to the 'best' coordinate system in an explicit form. Remember that, according to the results of Section 5.2, the form of the Lagrangian in this coordinate system can also be written down explicitly for any even $N$. Using formulas of (5.6), (2.10) and (6.30) we get the following form of the Lagrange function for system (6.15):

(6.31)
$$L(x, \dot{x}) = \sum_{i=1}^{N}\left(\sum_{n=1}^{N-2} C_n(x)\frac{\partial(C_N(x) + C_{n+1}(x))}{\partial x_i} + C_{N-1}(x)\frac{\partial(C_N(x) + C_1(x))}{\partial x_i}\right)\dot{x}_i - \sum_{n=1}^{N-1} C_n(x)$$

Consider as an example the simplest case with $N = 2$. It gives us:

(6.32) $\qquad C_1(x) = x_1 + x_2, \quad C_2(x) = (x_1^2 + x_2^2)/2$

and results in the Lagrangian

(6.33) $\qquad L(x, \dot{x}) = (x_1 + x_2)(x_1\dot{x}_1 + x_2\dot{x}_2) - x_1 - x_2$

Now apply to it the Lagrange equations. We have

(6.34) $\qquad \dfrac{d}{dt}\dfrac{\partial L}{\partial \dot{x}_1} = (2x_1 + x_2)\dot{x}_1 + x_1\dot{x}_2, \quad \dfrac{d}{dt}\dfrac{\partial L}{\partial \dot{x}_2} = (2x_2 + x_1)\dot{x}_2 + x_2\dot{x}_1$

and

(6.35) $\qquad \dfrac{\partial L}{\partial x_1} = (2x_1 + x_2)\dot{x}_1 + x_2\dot{x}_2 - 1, \quad \dfrac{\partial L}{\partial x_2} = (2x_2 + x_1)\dot{x}_2 + x_1\dot{x}_1 - 1$

This leads us to the equations

(6.36) $\qquad (x_1 - x_2)\dot{x}_2 = -1, \quad (x_2 - x_1)\dot{x}_1 = -1$

which obviously coincide with the original equations (6.15) with $N = 2$ after division by the factor $\Phi_{12} = x_1 - x_2$ (see formula (1.12)).

Similarly, one can prove by explicit calculations that the Lagrange equations for $N = 4$ case also result in the original equation.

# 7. Summary

Here is a short summary of the obtained results:

- Any even-dimensional system of first-order dynamical equations

$$\dot{x}_i = R_i(x), \quad i = 1, \dots, 2M$$

  with no explicit dependence on time and not necessarily corresponding to any 'physical' or 'conservative physical' system can be derived from a certain Lagrange function $L(x, \dot{x})$ having no explicit dependence on time and no additional degrees of freedom. This Lagrange function is linear in system velocities:

$$L(x, \dot{x}) = \sum_{k=1}^{2M} P_k(x) \dot{x}_k - H(x)$$

By analogy with the 'physical' case, we will call its free term $H(x)$ – the 'hamiltonian', and the coefficients at the velocities $P_i(x)$ – the 'momenta'.

- The Lagrange equations

$$\frac{d}{dt} \frac{\partial L(x, \dot{x})}{\partial \dot{x}_i} = \frac{\partial L(x, \dot{x})}{\partial x_i}, \quad i = 1, \ldots, 2M$$

for the above function take the form

$$\sum_{k=1}^{2M} F_{ik}(x) \dot{x}_k = \frac{\partial H(x)}{\partial x_i}, \quad i = 1, \ldots, 2M$$

where

$$F_{ik}(x) = \frac{\partial P_k(x)}{\partial x_i} - \frac{\partial P_i(x)}{\partial x_k}$$

is a certain anti-symmetric tensor explicitly expressible via the $2M - 1$ integrals of motion of the dynamical system and one additional time-like integral (for details see Section 2).

- In cases when tensor $F_{ik}(x)$ is invertible, the Lagrange equations can be reduced to the form (0.1) with

$$R_i(x) = \sum_{k=1}^{2M} F_{ik}^{-1}(x) \frac{\partial H(x)}{\partial x_k}$$

- It is natural to ask the following question: How to construct the functions $H(x)$ and $P_i(x)$ in a way ensuring the invertibility of the tensor $F_{ik}(x)$ provided that the functions $R_i(x)$ – the right-hand sides of equation (0.1) – are given? The corresponding construction is performed in Section 2. It is shown that the free term $H(x)$ can be explicitly constructed from the standard integrals of motion $C_1(x), \ldots, C_{2M-1}(x)$ of system (0.1) – i.e., functions not changing in time as system evolves. The coefficients at the velocities $P_i(x)$ are also functions of coordinates only constructible from the same integrals of motion plus one additional function of coordinates $C_{2M}(x)$ behaving as time as system evolves. For details of such a construction see Section 2.

- The free term $H(x)$ is a Noetherian integral of motion induced by time-symmetry of the Lagrange function:

$$H(x) = \sum_{k=1}^{2M} \frac{\partial L(x, \dot{x})}{\partial \dot{x}_k} \dot{x}_k - L(x, \dot{x})$$

It is a natural generalization of the classical Hamiltonian function for non-physical and/or non-conservative systems. The coefficients at the velocities $P_k(x)$ are natural generalizations of classical momenta:

$$P_k(x) = \frac{\partial L(x, \dot{x})}{\partial \dot{x}_k}$$

- The momenta are defined up to the gradient transform

$$P_k(x) \rightarrow P_k(x) + \frac{\partial \Phi(x)}{\partial x_k}$$

equivalent to adding a total time-derivative of $\Phi(x)$ to the Lagrangian:

$$L(x, \dot{x}) \rightarrow L(x, \dot{x}) + \frac{d\Phi(x)}{dt}$$

This transform does not affect the form of the equations of motion leaving the antisymmetric tensor $F_{ik}(x)$ invariant. This makes this transform similar to the 'gauge transform' in classical electrodynamics with the following obvious analogies:

| $P_i(x)$ | Vector-potential |
| --- | --- |
| $F_{ik}(x)$ | Tensor of Electromagnetic Field |
| $\Phi(x)$ | Gauge Scalar |

- Despite the fact that the 'gauge transform' does not formally affect the form of equations of motion, it is far from being an innocent operation from the standpoint of the variation formalism. The point is that in case of Lagrangians linear in system velocities not all gauges are consistent with the stationary action principle and not necessarily allow the derivability of equations of motion within the standard variation approach. To make such a derivation possible the 'gauge' must be fixed in special ways.

- It turns out that the corresponding gauge-fixing condition has the form of the classical stationary Hamilton-Jacobi equation:

$$\mathcal{H}\left(\frac{\partial \Phi(x)}{\partial x_1}, \ldots, \frac{\partial \Phi(x)}{\partial x_{2M}}\right) = const$$

whose left-hand side is determined by the function $H(x)$ rewritten in terms of 'momenta' $p$. Formally this means that

$$\mathcal{H}(p_1, \ldots, p_{2M}) = H(x_1(p), \ldots, x_{2M}(p))$$

where $x_i(p)$ are solutions of the system of equations: $p_i = P_i(x)$.

- Introduction of gauge-based corrections to the Lagrange function allows one to interpret the gauge $\Phi(x)$ as action integral taken along the true trajectory of the system and considered as function of the upper limit of integration. Otherwise it would be impossible because of the null-property of the Lagrangian according to which the value of the Lagrangian on all the true trajectories of system is zero.

- The original equations of motion can be rewritten in a more compact and universal form:

$$\dot{x}_i = \{x_i, \; H(x)\} \quad i = 1, \ldots, 2M$$

  after introducing a special analog of Poisson brackets $\{\ldots, \ldots\}$. Exactly as in the classical case, the conservation laws can be expressed as condition of vanishing the Poisson bracket with the Hamiltonian. Many properties of the classical Poisson brackets (like anti-symmetry, multiplication property and Jacobi identity) are preserved. Although there are natural analogs of canonical relations, we couldn't find any meaningfully defined analogs of canonical transforms preserving these relations.

- The most general structure-preserving transform is the general coordinate transform. One of its particular cases induced by the evolution of the system does not conserve the volume of the 'phase' space defined as $dV = dx_1 \ldots dx_{2M}$ in our case. We introduce a corrected volume $dV' = J(x)dV$ conserved during the evolution of the system. This allows one to formulate an analog of the Liouville theorem. The relationship between the volume and information conservation is discussed.